\documentstyle[12pt,psfig]{article} 
\newcommand{\Li}{{\rm Li}}
\newcommand{\dd}{{\rm d}}
\def\lnomega{\ln(\omega)}
\def\lnomegad{\ln^2(\omega)}
\def\lndelta{\ln(1-\omega)}
\newcommand{\be}{\begin{eqnarray}}
\newcommand{\ee}{\end{eqnarray}}
\newcommand{\ep}{\epsilon}
\newcommand{\om}{\omega}

\begin{document}

\rightline{TTP97-08}
\rightline{hep-ph/9703277}
\rightline{March 1997}
\bigskip
\begin{center}
{\Large {\bf Two-loop QCD Corrections to $b \to c$ Transitions 
 at Zero Recoil:}}\\[2mm]
{\Large{\bf Analytical Results}}
\end{center}
\vspace{0.2cm}
\smallskip
\begin{center}
{\large{\sc Andrzej Czarnecki}}
\hspace{0.1cm}
and 
\hspace{0.1cm}
{\large{\sc Kirill~Melnikov}}
\\
\vspace{0.7cm}
{\sl Institut f\"{u}r Theoretische Teilchenphysik}\\
{\sl Universit\"{a}t Karlsruhe}\\
{\sl D--76128 Karlsruhe, Germany}\\[2mm]
{e-mail: \sl ac,melnikov@ttpux8.physik.uni-karlsruhe.de}\\
\vspace{1.8cm}
{\large{\bf Abstract}}\\
\end{center}
\vspace{0.1cm}
We present analytical results for the $O(\alpha _s ^2)$ contributions to
the functions $\eta _A$ and $\eta _V$ which parameterize QCD
corrections to semileptonic $b \to c$ transitions at zero recoil.
Previously obtained approximate
results are confirmed.
The methods of computing the relevant two-loop diagrams with two mass
scales are discussed in some detail.  
\newpage

\section{Introduction}
The studies of the semileptonic decays of the $b$--quark provide the
best opportunity to determine $|V_{cb}|$, a parameter of the CKM
matrix and a fundamental input parameter in the Standard
Model. Currently, two different methods of extracting the $|V_{cb}|$
 from the experimental data are used. One of them is based on the
exclusive decays $B \to D^{\star}l\nu _l$.
In this method the recoil spectrum of the $B$ meson decay is measured:
\begin{eqnarray}
{{\rm d} \Gamma(B\to D^* l\bar \nu) \over {\rm d}w}
= f(m_B,m_{D^*},w) |V_{cb}|^2 {\cal F}^2 (w)
\left( 1+ {\alpha\over \pi}\ln {M_Z\over m_B} \right).
\label{eq:spectrum}
\end{eqnarray}
In this equation $w$ is the product of the four velocities of the $B$
and $D^{\star}$ mesons and $f$ is a known function of the masses of
the observed particles.

In the limit when the $b$- and $c$- quark masses are considered to be
large in comparison with $\Lambda _{QCD}$, the Heavy Quark Effective
Theory \cite{Isgur89,Isgur90,Eichten90,SV88} 
can be used to determine the nonperturbative corrections
responsible for deviations of ${\cal F}(1)$ from unity
in a model independent way. For this purpose, small velocity sum rules
provide a reliable theoretical framework \cite{Shifman94}.  

 From the experimental point of view, the zero recoil point 
is not directly accessible
due to the vanishing phase space. Therefore, one measures
$|V_{cb}|^2~{\cal F}^2(w)$ 
for $w \ne 1$, extrapolates it to $w=1$ and uses theoretical
predictions for ${\cal F}(1)$ to extract the value of $|V_{cb}|$.

Apart from the abovementioned nonperturbative effects, ${\cal F}(1)$
receives also perturbative QCD corrections.  The Lorentz structure of
the $b \to c$ decay vertex is $\Gamma _\mu = \gamma _\mu (1-\gamma
_5)$.  We describe the modifications of the axial and vector parts at
zero recoil by two functions:
\be
\gamma _\mu \to \eta _V \gamma _\mu,~~~
\gamma _\mu \gamma _5 \to \eta _A \gamma _\mu \gamma _5.
\ee

Both functions $\eta _A$ and $\eta _V$ can be expanded in power series
in the strong coupling constant:
\be
\eta _{A,V} = 1 + \frac {\alpha _s}{\pi}C_F\eta _{A,V} ^{(1)} +
\left(\frac {\alpha _s}{\pi} \right)^2 C_F\eta _{A,V} ^{(2)}
+O(\alpha _s ^3). 
\ee
The $O(\alpha _s)$ effects have been known for a long time
\cite{Paschalis83,Close84,SV88}.
The $O(\alpha _s ^2)$ corrections to zero recoil form factors have
been subject of controversy over several years. The need to
perform complete two-loop calculation has been emphasized
by many authors. The problem was solved in ref.~\cite {zerorecoil},
where $O(\alpha _s ^2)$ were calculated in the form of a series
in $\delta \equiv 1-m_c/m_b$. The value of  $O(\alpha _s ^2)$ turned out to
be relatively small and therefore the important source of the
theoretical uncertainty in the $|V_{cb}|$ determination from exclusive
decays has been removed.

The purpose of the present paper is to present  analytical results
for the  $O(\alpha _s ^2)$ corrections to 
two functions $\eta _A$ and $\eta _V$ and to discus the methods
which have been used for their evaluation.  The Feynman diagrams which
have to be computed are shown if Fig.~1.

\begin{figure} 
\hspace*{0mm}
\begin{minipage}{16.cm}
\vspace*{5mm}
\[
\mbox{
\begin{tabular}{ccc}
\psfig{figure=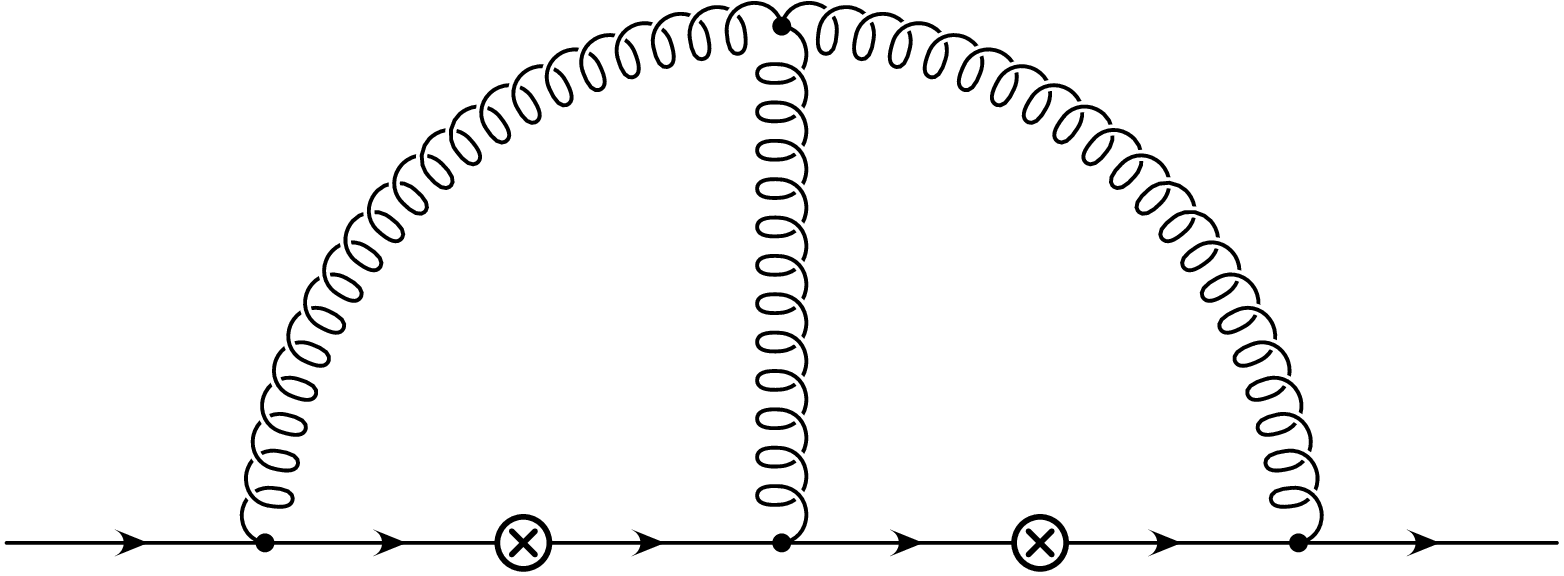,width=23mm,bbllx=210pt,bblly=410pt,%
bburx=630pt,bbury=550pt} 
&\hspace*{19mm}
\psfig{figure=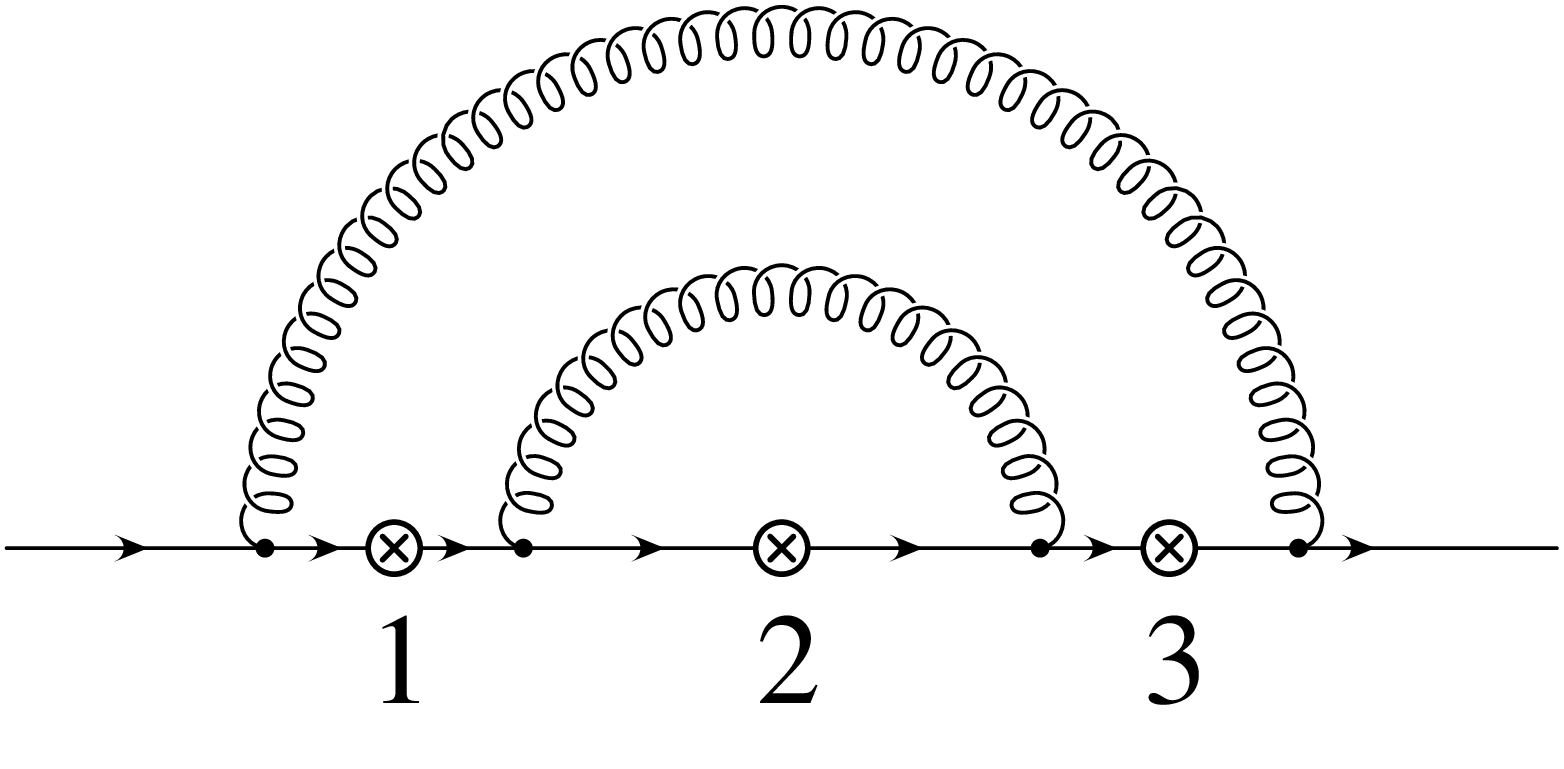,width=23mm,bbllx=210pt,bblly=410pt,%
bburx=630pt,bbury=550pt}
&\hspace*{19mm}
\psfig{figure=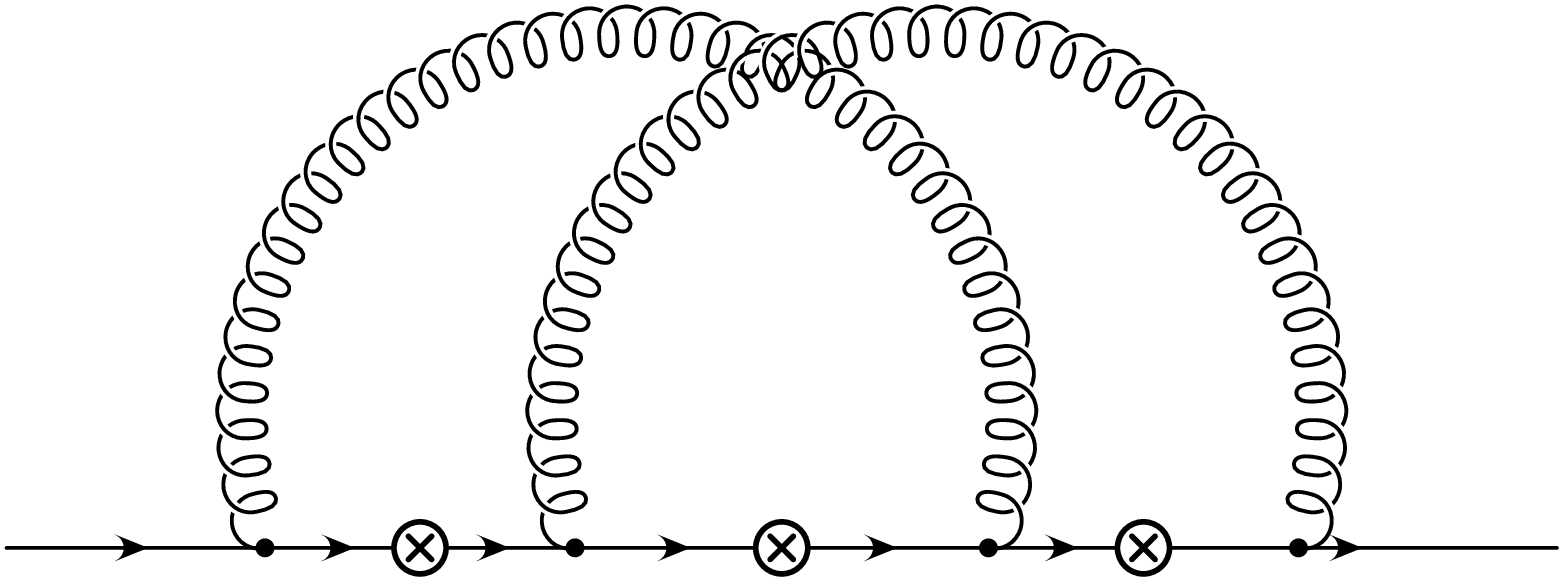,width=23mm,bbllx=210pt,bblly=410pt,%
bburx=630pt,bbury=550pt}
\\[5mm]
\hspace*{-13mm}(a) & \hspace*{7mm}(b) & \hspace*{7mm}(c) 
\\[10mm]
\psfig{figure=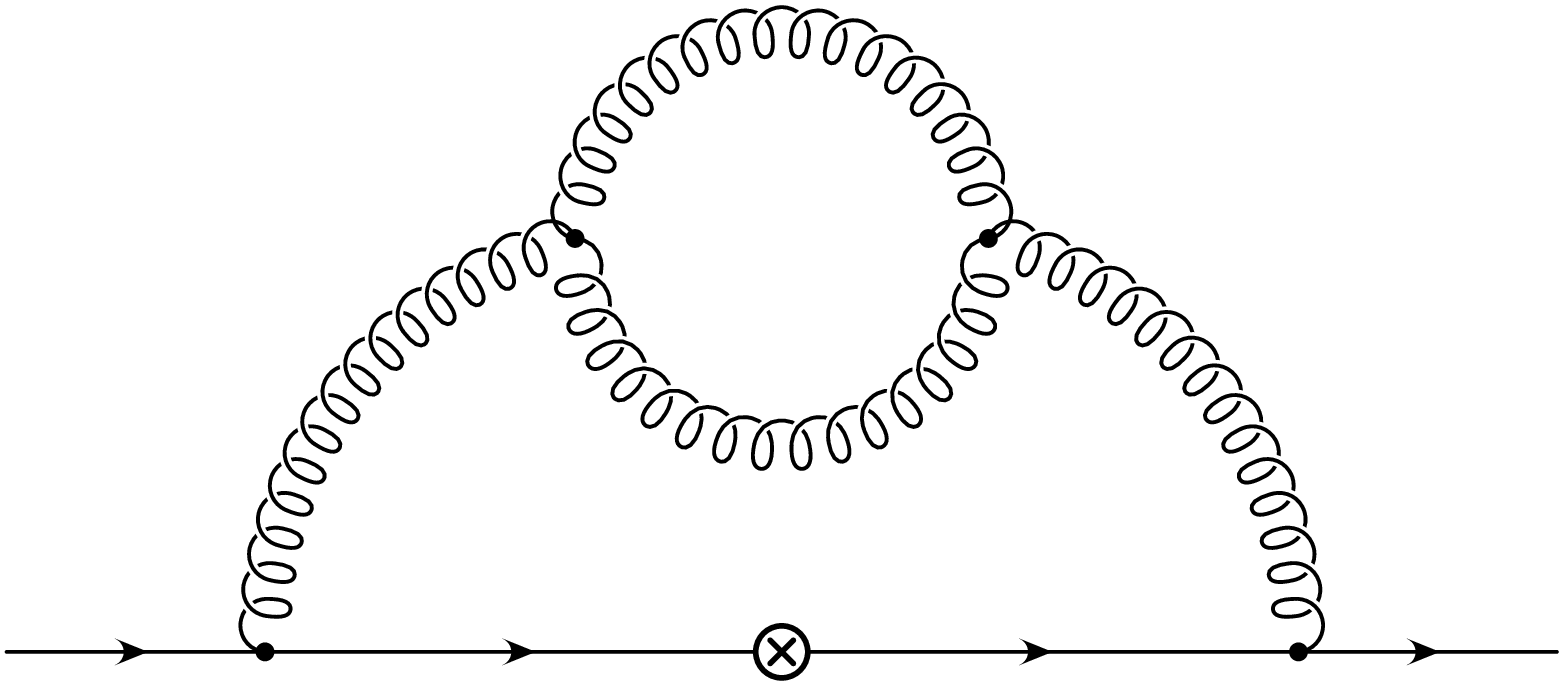,width=23mm,bbllx=210pt,bblly=410pt,%
bburx=630pt,bbury=550pt} 
&\hspace*{19mm}
\psfig{figure=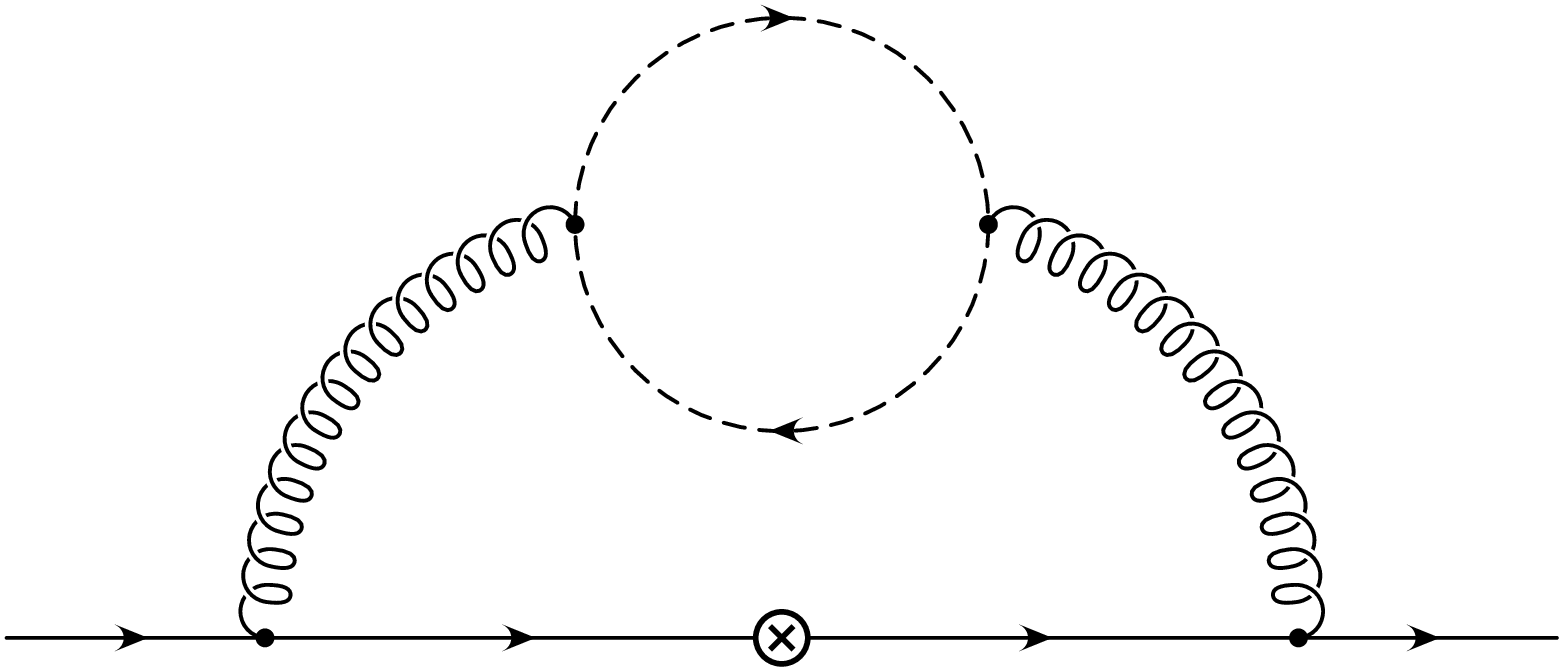,width=23mm,bbllx=210pt,bblly=410pt,%
bburx=630pt,bbury=550pt}
&\hspace*{19mm}
\psfig{figure=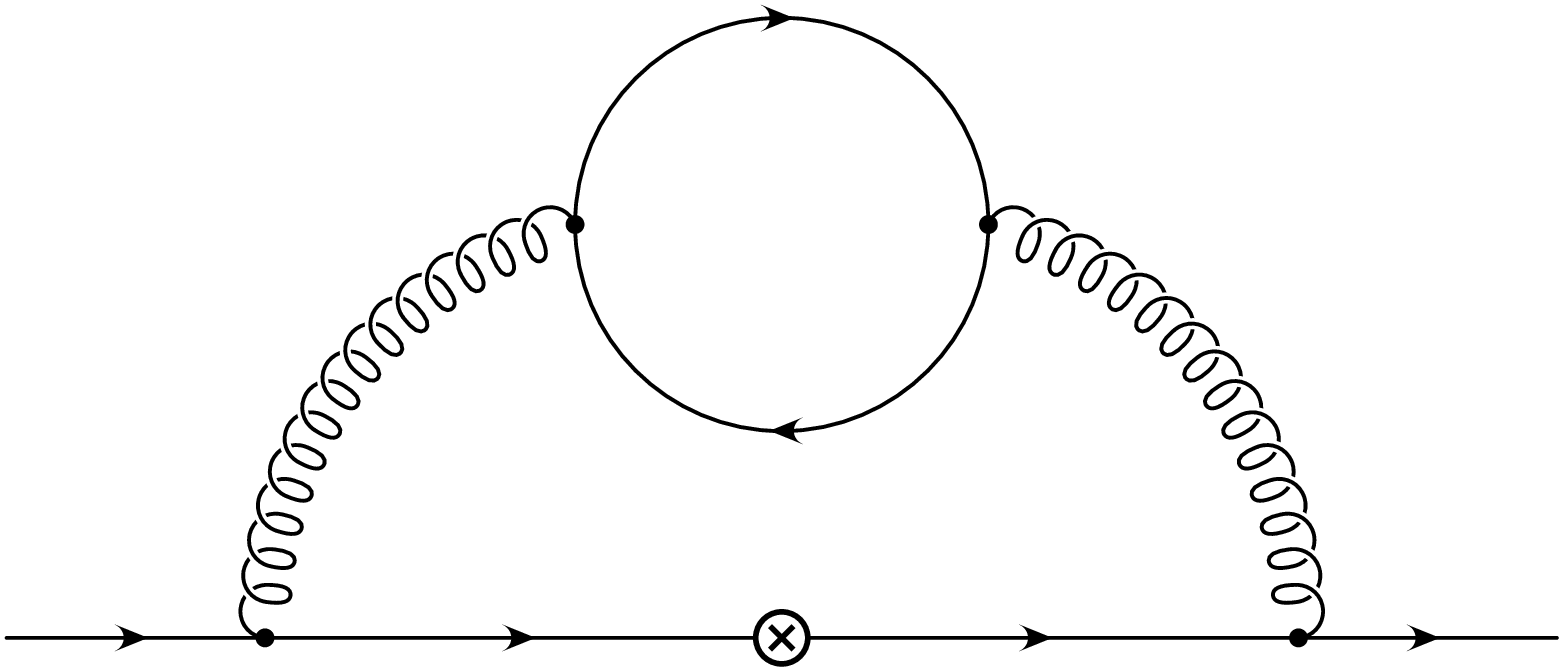,width=23mm,bbllx=210pt,bblly=410pt,%
bburx=630pt,bbury=550pt}
\\[5mm]
\hspace*{-13mm}(d) & \hspace*{7mm}(e) & \hspace*{7mm}(f) 
\end{tabular}}
\]
\end{minipage}
\caption{Two-loop QCD corrections to the $b\to c$ transitions at zero
recoil. Symbols $\otimes$ mark places where a virtual $W$ boson
can possibly couple to the quark line. } 
\label{fig:twoloop}
\end{figure}

There are several motivations for taking up this issue again.
First we want to demonstrate that with the current techniques for doing
multiloop calculations the calculation of  $O(\alpha _s ^2)$
correction to  $\eta _{A,V}$ is feasible and even
relatively simple, if  organized  properly.
Second, having a complete formula one can study it in the whole range
of  $m_c/m_b$, while the previously available expansion is not valid
for $m_c\ll m_b$.
Finally, it is worth reminding that there are almost no results on
the decays of  fermions neither in QED nor in QCD at the two loop level.
As a long standing example, the two loop QED corrections to the muon
decay are not known. Being a challenging theoretical problem, the
value of $O(\alpha ^2)$ corrections to muon decay does not 
influence precision  electroweak physics at present. However,
this situation may 
change if the electroweak parameters, such as
$\sin^2\theta_W$, are measured with higher accuracy, 
e.g.~at the future linear collider. 
In view of the anticipated progress in precision measurements, 
accumulating as much information as possible 
about higher order effects in the fermion decays is warranted.

The feasibility of the analytical calculation at zero recoil is mainly
connected with the particular kinematical configuration. 
We denote the momentum of
the $b$--quark in the initial state by $p$ with the on-shell condition
$p^2 = m_b^2$. At zero recoil, 
the $c$--quark in the final state has the momentum
$\omega p$, where $\om = m_c/m_b$. The virtual $W$ boson, which
induces the semileptonic transition,  carries
off the momentum $(1-\om)p$. Hence, the zero recoil
kinematics dictates that the momenta of the particles in the final
state are parallel to the momentum of the particle in the initial state.

To put this in more simple terms, let us consider the zero recoil $b
\to c$ transition in the rest frame of the $b$ quark. In this case the
$c$--quark in the final state remains at rest and the energy
release $m_b - m_c$ is carried off  by the current which induces
this transition. From this picture it is evident that no real gluon
radiation can appear in the  zero recoil transition.

 From the technical point of view, the fact that all momenta are
parallel to each other  simplifies significantly the
calculations of the Feynman integrals. The Feynman integrals
which appear are similar to those
 studied in the context of the quark wave function and mass
renormalization constants calculation, except that here we have to
deal with two (rather than one) mass scales.  Below we shall
demonstrate that if this similarity is properly exploited, the
analytical calculation of the form factors at zero recoil turns out to
be feasible and even relatively simple.

The paper is organized as follows.  In section \ref{sec:basic} we
describe in some detail methods of computing two-loop integrals with
two mass scales needed in the zero recoil calculation.  We pay special
attention to the subtraction of infrared divergences so that the
difficult parts of the calculation can be done in 4 dimensions.  In
section \ref{sec:massive} the special case of diagrams with a massive
fermion loop inserted in a gluon propagator is considered.  Our final
result is presented and compared with the previous approximations in
section \ref{sec:results}.

\section{Basic scalar integrals}
\label{sec:basic}

To extract a contribution of a given Feynman diagram to $\eta _{A,V}$
we multiply it by $\bar u _c \gamma _{\mu} u _b$ or 
$\bar u _c \gamma _{\mu} \gamma _5 u _b$ and average over 
fermion polarizations. In this way one gets scalar products 
of four momenta in the numerator which one expresses through the 
available denominator structures. This leads to a
reduction of the number of the propagators for a given Feynman diagram.
After this step the  remaining Feynman integrals have to be calculated.

Before demonstrating various types of integrals we introduce some
notations. We denote the loop momenta by $k$ and $l$ and work in the
Euclidean space.  There are nine possible propagators
which appear in the Feynman integrals is this calculation:
\be
S_1 = k^2,~~S_2=l^2,~~S_3 = (k-l)^2,
\ee
$$
S_4 = k^2+2pk,~~~S_5 = l^2+2pl,~~~S_6=(k+l)^2+2p(k+l),
$$
$$
S_7 = k^2+2\om pk,~~~S_8 = l^2+2\om pl,~~~S_9=(k+l)^2+2\om p(k+l).
$$

An arbitrary Feynman integral constructed from these objects will be
denoted as
\be
D(\alpha _1, \alpha _2, \alpha _3, \alpha _4, \alpha _5, \alpha _6,
\alpha _7, \alpha _8, \alpha _9|p,\om) = 
\int \frac {[\dd k][\dd l]}{S_1^{\alpha _1}S_2^{\alpha _2}S_3^{\alpha _3}
S_4^{\alpha _4}S_5^{\alpha _5}S_6^{\alpha _6}S_7^{\alpha _7}S_8^{\alpha _8}
S_9^{\alpha _9}}
\ee
with
\be
[\dd k]=\frac {d^Dk}{\pi^{D/2}},~~~~D=4-2\ep.
\ee

If the last two arguments of $D(....|Q,x)$ are shown explicitly they
denote objects which should replace $p$ and $\omega$ in $S_i$.
Otherwise it is assumed that $Q=p$ and $x=\om$.  We use $m_b$ as a
unit of mass. The proper dimension of the integrals can be easily
restored, if needed.

For the presentation of our results for the 
relevant Feynman integrals it is
convenient to introduce two auxiliary functions:
\be
R_1 &=&
 \frac {\pi ^2}{6}-\Li _2 (1-\om)= \Li _2(\om)+\ln (1-\om) \ln(\om),
\nonumber \\
R_2 &=& \Li _2(-\om)+\ln (1+\om) \ln(\om).
\label{eq:r12}
\ee
If the argument of $R_{1,2}$ is given explicitly, it replaces $\om$ in
the above definitions.

\subsection{Scalar integrals with six propagators}

The highest number of propagators in Feynman integrals 
in the zero recoil calculation is six. Both planar (like in Fig.~1a,b)
and
non--planar (Fig.~1c) integrals  with six propagators appear.

\subsubsection*{Planar diagrams with six propagators} 

Let us consider the integral $D(1,0,1,1,1,0,1,1,0)$:
\be
D(1,0,1,1,1,0,1,1,0) = \int \frac {[\dd k][\dd l]}{k^2~(l-k)^2 (k^2+2pk)
(l^2+2pl)(k^2+2\om pk)(l^2+2 \om pl)}.
\ee

The idea which permits the calculation of this integral is the
following.  Consider the product of $S_4$ and $S_7$ or of $S_5$ and
$S_8$ in the integrand.  For both it is possible to perform partial
fractioning, like
\be
\frac {1}{(k^2+2pk)(k^2+2\om pk)} = \frac {1}{1-\om} \frac
{1}{k^2}\left( \frac {1}{k^2+2pk}-\frac {\om}{k^2+2\om pk} \right).
\ee
For $D(1,0,1,1,1,0,1,1,0)$ we therefore write
\be
{1\over S_4S_7S_5S_8}
 = \frac {1}{(1-\om)^2}{1\over S_1S_2}
 \left({1\over S_4} - {\om\over S_7}\right)
 \left({1\over S_5} - {\om\over S_8}\right)
\ee
and, accordingly,
\be
D(1,0,1,1,1,0,1,1,0) &=& \frac {1}{(1-\om)^2} 
\Bigg [ D(2,1,1,1,1,0,0,0,0)+
\om^2 D(2,1,1,0,0,0,1,1,0)
\nonumber \\ &&
-\om D(2,1,1,1,0,0,0,1,0)-\om D(2,1,1,0,1,0,1,0,0) \Bigg ]. 
\label {reduced}
\ee

We notice that the first two $D$ structures in the square brackets are
two--loop two--point on--shell integrals with a single mass scale.
The only difference between them is in the incoming momenta and in the
masses of the particles inside the loop. An algorithm to compute an
arbitrary integral of this type is known \cite{gbgs90,GrayPhD,bro91a}
(the underlying idea is the integration by parts method \cite{che81}).
A large number of such integrals had to be computed in the approximate
approach to the present problem \cite {zerorecoil}.  For that purpose
a new implementation of the recurrence algorithm was necessary.  The
ability to compute such integrals determines our strategy also in the
present case.

Apart from the known single scale integrals, eq.~(\ref {reduced})
contains two unknown integrals. 
Let us consider one of them as an example:
$$
D(2,1,1,1,0,0,0,1,0)= 
\int \frac {[\dd k][\dd l]}{(k^2)^2l^2(l-k)^2(k^2+2pk)(l^2+2\om pl)}.
$$

The price we have to pay for partial fractioning 
is that the integrals we obtain
have stronger infrared divergences than the original ones. 
The integration in
$D$--dimensions is much more complicated than if the number of
space--time dimensions could be equated to 4 from the start.
In circumventing the problem the following trick proved to be useful.
We introduce a Feynman parameter to combine two propagators in the
integrand of $D(2,1,1,1,0,0,0,1,0)$:
$$
\frac {1}{(k^2)^2(k^2+2pk)} = 2\int \limits _{0}^{1} 
\frac {\dd t~(1-t)}{(k^2+2pkt)^3}.
$$
Introducing the variable $t=\om \mu$ one gets
$$
\int \limits _{0}^{1} 
\frac {\dd t~(1-t)}{(k^2+2pkt)^3} = \om \int \limits _{0}^{1/\om}
\frac {\dd \mu (1-\om \mu)}{(k^2+2pk\om \mu)^3}.
$$
Splitting the integration over $\mu$ into two parts: 
 from $0$ to $1$ and from $1$ to $1/\om$,
we get the following representation for $D(2,1,1,1,0,0,0,1,0)$:
\be
D(2,1,1,1,0,0,0,1,0) &=& 
\om D(2,1,1,0,0,0,1,1,0) 
+ \om (1-\om) D(1,1,1,0,0,0,2,1,0)
\nonumber \\ &&
 + \om (1-\om)^2 D(0,1,1,0,0,0,3,1,0)
+\Delta, \label {reduced1}
\ee
where $\Delta$ is:
$$
\Delta = 2\om \int \limits _{1}^{1/\om} \dd \mu (1-\om \mu)\left[
D(0,1,1,0,1,0,3,0,0|p\om,\mu)
-\frac {1}{\mu ^3} D(0,1,1,0,1,0,3,0,0|p\om,1) \right].
$$

The advantage of this representation is that all the $D$
functions, listed in  eq.~(\ref {reduced1}) are now of the single scale
type and their values are available.
On the other hand, the part of the result denoted by $\Delta$
is finite and its calculation can be performed in
four dimensions. 

Let us stress this point once again. One of the obstacles in
calculating  the zero recoil form factors is the fact that 
individual Feynman integrals have infrared divergences, which
should disappear in the sum. 
It is the appearance of these
divergences, which makes the calculation tough.  We have demonstrated
above that in the  zero recoil calculation
one can  extract the infrared divergences in the form of the 
on--shell  integrals which are  known.

For the sake of illustration, we write down 
the following finite result:
\be
D(0,1,1,0,1,0,3,0,0|p\om,\mu)-{1\over \mu^3} 
D(0,1,1,0,1,0,3,0,0|p\om,1)=
\frac {1}{2 \om ^4} \frac {1-\mu+\ln \mu}{(1-\mu) \mu ^3}.
\ee
The simplicity of this result  leaves no doubts
that the remaining integration over $\mu$ can easily be performed. 
We thus arrive at the  result for the
planar diagram with six propagators:
\be
\lefteqn{D(1,1,0,1,1,0,1,1,0) }
\nonumber \\  &&
=  - \frac {1}{\om ^2}\left[ \frac {12\ep^2+2\ep-1}{8\ep ^2} 
 -{1+\om\over 1-\om}\Li_2(1-\om)
 -\frac {(2+\om) \ln^2 (\om)}{2(1- \om )}  
+\frac {(1-2\ep)\ln (\om)}{2\ep(1-\om)} 
    - \frac {13\pi ^2}{48} \right].
\nonumber \\
\ee
This is the only planar diagram with six different 
propagators which is necessary for the present calculation.

\subsubsection*{Non-planar diagrams  with six propagators} 

Another integral with 6 propagators we need 
is $D(1,1,0,1,1,1,0,1,0)$.
Performing partial fractioning we get
$$
D(1,1,0,1,1,1,0,1,0)=\frac {1}{1-\om}D(1,2,0,1,1,1,0,0,0)-
\frac {\om}{1-\om}D(1,2,0,1,0,1,0,1,0).
$$
The first of these integrals is of the single scale type.
To calculate the second one, we use the same trick as described above and 
get the following representation:
\be
\lefteqn{D(1,2,0,1,0,1,0,1,0)=\frac {1}{\om}D(1,2,0,1,1,1,0,0,0)}
\nonumber \\&&
-\frac {1-\om}{\om ^2}D(1,1,0,1,2,1,0,0,0)-\frac {2}{\om}
\int \limits _{\om}^{1} \frac {\dd t}{t} \left( 1-\frac {t}{\om} \right)
D(0,1,0,3,0,0,0,1,1|t ^{-1})  .
\ee
The first two integrals in this expression are of the single scale type. 
The expression for the Feynman diagram under the $t$
integration will be presented later (eq.~\ref{eq:d3}). 
Direct integration gives: 
\be
\int \limits _{\om}^{1} \frac {\dd t}{t} \left( 1-\frac {t}{\om}\right)
\!\!\! && \!\!\!
D(0,1,0,3,0,0,0,1,1|t^{-1})=
 -\frac {(1-3\om)^2}{16\om ^2}R_1 
+\frac {(1+\om)^2}{16\om^2}R_2
\nonumber \\ &&
-\frac {1-\om}{4\om }\ln^2(\om)-\frac {5}{8\om}\ln(\om)-
\frac {7(1-\om)}{8\om} - \frac {\pi ^2(1-4\om)}{48\om } .
\ee
 From this an expression for $D(1,1,0,1,1,1,0,1,0)$ can be
constructed.

The diagram $D(1,1,0,1,0,1,0,1,1)$
represents a more challenging problem. We  note that it
has an overall infrared divergence and cannot be calculated in $D=4$.
To compute it we introduce the operator 
$$
\hat L = p_\mu \frac {\partial} {\partial p_\mu}.
$$
For dimensional reasons the integral we are interested in is
proportional to $(p^2)^{-2-2\ep}$. By applying the operator
$\hat L$ to it one finds
\be
\hat L D(1,1,0,1,0,1,0,1,1) = -(4+4\ep)~D(1,1,0,1,0,1,0,1,1).
\ee
By applying the same operator to the integrand we get
\be
-(1+4\ep) 
\!\!\!\! &&\!\!\!\! 
D(1,1,0,1,0,1,0,1,1) =D(0,1,0,2,0,1,0,1,1)+D(1,0,0,1,0,1,0,2,1)
\nonumber \\ &&
+\frac {1}{1-\om}D(1,1,0,1,0,2,0,1,0) - 
\frac {\om}{1-\om} D(1,1,0,1,0,0,0,1,2).
\ee
It is useful to note that Feynman  integrals in the
above equation satisfy the following symmetry relations
$$
D(1,0,0,1,0,1,0,2,1)=
\omega^{-4-4\ep}~D(0,1,0,2,0,1,0,1,1|\om^{-1}), 
$$
$$
D(1,1,0,1,0,0,0,1,2) = 
\omega^{-4-4\ep}~D(1,1,0,1,0,2,0,1,0|\om^{-1}).
$$
Therefore, it is enough to know two of them.

We present  the calculation of $D(0,1,0,2,0,1,0,1,1)$
which turns out to be the most complicated integral.
Applying the operator $\hat L$ to it
once more, we get the following identity:
\be
\lefteqn{
-2D(0,1,0,2,0,1,0,1,1) = \frac {2}{1-\om}D(0,1,0,3,0,1,0,1,0)}
\nonumber \\ &&
-\frac {2}{1-\om}D(0,1,0,3,0,0,0,1,1)+
\frac {2}{\om}D(0,1,0,3,0,1,0,0,1)
-\frac {2}{\om}D(0,0,0,3,0,1,0,1,1)
\nonumber \\ &&
+\frac {1}{1-\om}D(0,1,0,2,0,2,0,1,0)
+D(0,0,0,2,0,1,0,2,1) -\frac {\om}{1-\om}D(0,1,0,2,0,0,0,1,2).
\nonumber \\ 
\ee

All Feynman integrals in this equation are finite and relatively
simple to calculate directly.
The results for individual Feynman integrals read:
\be
D(0,1,0,3,0,0,0,1,1) &=& \frac {1}{2\om}\left[
\frac {1}{4}\Big (R_2-R_1  \Big )+\frac
{\om \ln(\om)}{2(1-\om)^2} +\frac {1}{2(1-\om)}+\frac {\pi^2}{8} \right],
\label{eq:d3}
\\
D(0,1,0,2,0,0,0,1,2) &=& \frac {1}{\om ^2}\left[
\frac {1}{4}\Big (R_2-R_1  \Big )-\frac
{\om \ln (\om )}{2(1-\om)^2} -\frac {1}{2(1-\om)}+\frac {\pi^2}{8} \right],
\nonumber\\
D(0,1,0,3,0,1,0,0,1)&=& - \frac {1-\om+ \ln(\om)}{2 (1-\om)^2},
\nonumber\\
D(0,0,0,3,0,1,0,1,1)&=& \frac {1+\om}{\om(1-\om)}\left[\frac {1}{8} 
\Big (R_1 -R_2 \Big )-\frac{\om \ln (\om)}{4(1+\om)^2}-\frac
{\pi ^2}{16}\frac {\om}{1+\om} \right],
\nonumber\\
D(0,0,0,2,0,1,0,2,1)&=&-\frac {1+\om}{\om^2(1-\om)}\left[\frac {1}{4} 
\Big (R_2-R_1  \Big )-\frac{\om\ln(\om)}{2(1+\om)^2}+\frac
{\pi ^2}{8}\frac {\om}{1+\om} \right],
\nonumber
\ee
$$
2D(0,1,0,3,0,1,0,1,0)+ D(0,1,0,2,0,2,0,1,0)=
\frac {1}{2\om} \Big ( R_1 - R_2  \Big ).
$$
 
The last missing contribution for the scalar non--planar diagram with
six propagators is $D(1,1,0,1,0,2,0,1,0)$. This integral is infrared
divergent and therefore its direct evaluation is not convenient.
However, the trick described in the context of the
planar integral with six propagators is useful also here.
One finds the following representation:
$$
D(1,1,0,1,0,2,0,1,0)=\frac {1}{\om}\Bigg [ D(1,1,0,1,1,2,0,0,0)-
\int \limits_{\om}^{1}\frac {\dd t}{t^4} D(0,1,0,2,0,0,0,1,2| t^{-1})
\Bigg ].
$$
The result for the Feynman integral under the integration over $t$ is
presented in the the list of the integrals given above. The final
integration over $t$ is elementary.
As a result we find
\be
\lefteqn{D(1,1,0,1,0,1,0,1,1) }
\nonumber\\ &&
=~ - \frac {1}{4\om ^2\ep}
+ \frac {1}{4\om ^2(1-\om)}\Bigg [
2(1+\om)(R_1 -R_2)+4\ln(\om) +12(1-\om)-\pi ^2 \Bigg ].
\ee

\subsection{Scalar integrals with five propagators}

The scalar integrals with five propagators represent probably the most
difficult part of this work. The reason is that these integrals contain
trilogarithms, therefore evaluation of the corresponding Feynman
integrals is harder and the resulting expressions are rather
complicated.

\subsubsection*{Planar diagrams  with five propagators}

The most difficult  planar diagram is $D(1,1,1,1,0,0,0,1,0)$.
To calculate it we use the trick described in the previous section,
which allows us to extract on--shell divergent piece from this diagram.
The following representation  is useful:
$$
D(1,1,1,1,0,0,0,1,0) = \frac {1}{\om}\left[
D(1,1,1,1,1,0,0,0,0)-
 \int \limits_{\om}^{1}~\dd t~D(1,0,1,1,0,0,0,2,0|t)\right],
$$
where
$$
D(1,0,1,1,0,0,0,2,0|t)=-\frac {2}{t}R_1(t)-
\frac{\ln^2(t)}{1-t} +\frac {2\pi^2}{3t}.
$$

Performing the integration over $t$ one gets:
\be
D(1,1,1,1,0,0,0,1,0)&=& \frac {1}{\om}D(1,1,1,1,1,0,0,0,0)-
\frac { 6\Big (\Li _3(\om)-\zeta _3 \Big )}{\om} 
\nonumber \\
&&
 + \frac {1}{\om}  
\left[  4\ln (\om) \Li _2 (\om) +\ln^2(\om)\ln(1-\om)
+\frac {2\pi ^2}{3}\ln^2(\om) \right].
\ee

\subsubsection*{Non-planar diagrams  with five propagators}

This is the most difficult case.  Let us start our discussion with
$D(1,1,0,1,0,0,0,1,1)$.
We find that this diagram satisfies the following relation:
\be
D(1,1,0,1,0,0,0,1,1)=\frac {1}{\om ^2}H\left(\frac {1}{\om}\right),
\ee
where the function $H$ is given by
$$
H(\om)=\frac {1}{\om}\int \limits _{0}^{\om}
D(1,0,0,1,0,1,0,2,0|t)\dd t.
$$
For the expression under the integral we get the
following result:
\be
D(1,0,0,1,0,1,0,2,0|t) = -\frac {2(1+t)}{t(1-t)}R_2(t)+
\frac {\ln^2(t)}{1-t} - \frac {\pi ^2}{3(1-t)}.
\ee

This nice integral representation essentially solves the problem. The
subsequent integration over $t$ can be performed; however, the final
result is complicated and we relegate it to the appendix.
 For all practical purposes, like evaluation of the final
result as a function of $\om$, the integral representation provides a more
useful starting point than the exact formula.

There are two other integrals which are necessary. We will
discuss the calculation of one of them $D(0,1,0,1,0,1,0,1,1)$.
For the time being let us denote it as $F(\om)$.

It turns out to be convenient to derive a differential equation for
this function. Differentiating it with respect to $\omega $ one gets:
\be
\Bigg (-\frac {\dd}{\dd\om} - \frac {1-2\om}{\om(1-\om)} \Bigg )
F(\om) =J(\om), 
\ee
$$
J(\om)=-\frac {1}{\om}D(0,0,0,1,0,1,0,2,1)+
\frac {1}{1-\om}D(0,1,0,1,0,0,0,1,2).
$$
Writing $F(\om)=\rho (\om) G(\om)$ with
$$
\rho(\om) = \frac {1}{\om(1-\om)}
$$
one gets a new equation for $G(\om)$:
\be
\frac {\dd G(\om)}{\dd\om} = \om(1-\om) J(\om). \label {diffeq}
\ee

In order to determine the right hand side
of this equation we have to find $J(\om)$. 
For this we need two Feynman
integrals which can be computed directly
\be
D(0,1,0,1,0,0,0,1,2) &=& -\frac {2R_1}{\om} +
\frac {2(1+\om)}{\om(1-\om)}R_2-\frac {2\ln^2(\om)}{1-\om}
+\frac {\pi^2 (2-\om)}{3\om(1-\om)},
\nonumber \\
D(0,0,0,1,0,1,0,2,1) &=& \frac {2R_1}{\om} -\frac {2R_2}{1-\om}
+\frac {(1+2\om)\ln^2 (\om)}{1-\om ^2}-\frac {\pi^2(2-\om)}{3(1-\om^2)}.
\nonumber
\ee
and 
\be
J_1(\om) \equiv \om (1-\om)J(\om)=-\frac {2R_1 }{\om}+\frac {4R_2}{1-\om}
-\frac {(1+3\om)\ln^2(\om)}{1-\om^2}+
\frac {2\pi ^2(2-\om)}{3(1-\om^2)}.
\ee
With this input the differential
equation (\ref {diffeq}) leads to an
integral representation for $F(\om)$. 
$$
D(0,1,0,1,0,1,0,1,1)=\frac {1}{\om(1-\om)}\int \limits _{\om}^{1} \dd t 
J_1(t). 
$$
The analytical result can be found in the appendix.

\subsection{Feynman integrals with four propagators}

A lot of integrals with four propagators appear in the course of 
this calculation. A large fraction of them can be calculated directly.
In some cases however some special methods
appear to be useful.

Below we list some useful results.
For the diagram $D(0,0,0,1,0,1,0,1,1)$ we find the following:
$$
D(0,0,0,1,0,1,0,1,1)=
\frac {(1-\om ^{1-4\ep})~\Gamma (2\ep)~B(1-\ep,1-\ep)}
      {\ep~(1-3\ep)~(1-\om)}
-\frac {1}{1-\om}\Bigg (F_1(\om)-\om F_1\left(\frac {1}{\om}\right) \Bigg ),
$$
where
$$
F_1(\om)=-\frac {2(1-\om)^2R_1 }{\om}-\frac {2(1-\om)R_2}{\om}
+\frac {\om ^2\ln^2(\om )}{1+\om}+\frac {\pi^2(1+2\om^2)}{3(1+\om)}.
$$

For the diagram $D(1,0,0,1,0,1,0,1,0)$ we have:
\be
D(1,0,0,1,0,1,0,1,0) &=& \frac {\Gamma (2\ep)
    B(1-\ep,1-\ep)}{\ep(1-3\ep)}  
\nonumber \\ && \hspace*{-15mm}
-\frac {2(1+\om)}{\om}R_1 -\frac {2(1+\om)^2}{\om}R_2-
\frac {\om ^2\ln^2(\om)}{1-\om}-\frac {\pi^2(1+\om)}{3}.
\ee

Another type of diagrams we have to deal with are those
 with an internal massive bubble, for
example  $D(0,1,0,1,0,1,0,1,0)$. Similar integrals were
studied in \cite {gbgs90} in connection with the investigation of the 
quark wave function and mass renormalization constants. Methods
developed there permit a calculation of the integrals of
this type necessary in our case and will be useful also in case of
massive quark loop insertion in the gluon propagator (presented in the next
section). For example for
the diagram $D(0,1,0,1,0,1,0,1,0)$ we find
\be
D(0,1,0,1,0,1,0,1,0)&=&D(1,1,1,0,0,0,0,1,0)+
\frac {2(1-\om)^2R_1 }{\om ^2}
\nonumber \\ && 
+\frac {2(1+\om)^2R_2}{\om^2}
-2\ln^2(\om)-\frac {\pi ^2}{3}.
\ee

Other diagrams which appear in the course of this calculation, like
the diagrams with three propagators, can be reduced to the diagrams
with four propagators similar to the ones discussed above but with
the different structures in the numerator. All of these diagrams can
be studied in a similar manner and we do not present
them here in any detail.

The results presented above are sufficient to get the analytical
result for the major part of the diagrams involved in this
calculation. However, they are not sufficient for the diagrams
with the internal massive fermion loop. We shall discuss their
evaluation in the next section.

\section {Diagrams with a massive fermion loop}
\label{sec:massive}

The diagrams in Fig.~\ref{fig:twoloop}
with a massive $b$ or $c$ quark  loop insertion in the gluon
propagator
 represent a
somewhat 
special case.  It is  clear how to extract their divergences.
The ultraviolet
divergences  are canceled by wave function
renormalization. On the other hand, these diagrams give
infrared finite contributions if we use on--shell renormalization
for the coupling constant. Therefore, in order to deal with the
finite contributions from the very beginning one should study the
correction to the vertex together with the   
wave function renormalization and  renormalize the coupling
constant in a QED--like way. 

The contribution of the massive fermions to the gluon polarization
function $\Pi (k^2)$ can be written in a form of a dispersion 
integral (for the purpose of illustration we consider a particle in the 
internal loop to be $b$--quark) subtracted at $k^2 = 0$.
After that, the result appears as a convolution of a one-loop
correction evaluated with the gluon  of mass $\lambda $ with
the fermion spectral density. The following transformation
rule is valid:
\be
\frac {g_{\mu \nu}}{k^2} \to g_{\mu \nu} \frac {\alpha _s C_FT_R}{3\pi} 
\int \limits _{4m_b^2}^{\infty} \frac {d\lambda ^2}{\lambda ^2
(k^2-\lambda ^2)}\Bigg (1+\frac {2m_b^2}{\lambda ^2}\Bigg ) 
\sqrt {1-\frac {4m_b^2}{\lambda ^2}}.
\ee

To proceed further, we  first calculate
the one-loop form factors $\eta _{A,V}$ with a massive gluon. 
Afterwards the result must be integrated over the mass of the
gluon with the fermion spectral density:
\be
\eta _{A,V}(b) \sim \int \limits _{4m_b^2}^{\infty} 
\frac {d\lambda ^2}{\lambda ^2}\Bigg (1+\frac {2m_b^2}{\lambda ^2}\Bigg ) 
\sqrt {1-\frac {4m_b^2}{\lambda ^2}}~\eta _{A,V}^{(1,\lambda)}.
\label {gluonmassint}
\ee
Here $\eta_{A,V}(b)$ denotes the contributions of diagrams with a
$b$ quark loop inserted in the gluon propagator 
and  $\eta _{A,V}^{(1,\lambda)}$  denotes the one--loop form
factors calculated with a massive gluon propagator $1/(k^2-\lambda ^2)$.

It is advantageous to write the one--loop result leaving the
integration over the last of the Feynman parameters untouched. Namely,
both one--loop form factors calculated with the finite mass
of the gluon $\eta _{A,V}^{(1,\lambda)}$ can be written as:
\be
\eta _{A,V} ^{(1,\lambda)}=\int \limits _{0}^{1} d\beta 
\Bigg (\frac {P_1^{A,V}(\beta,\om)}{\lambda ^2 \beta + (1-\beta)^2} -
 \frac {P_2^{A,V}(\beta,\om)}{\lambda ^2 \beta +\omega ^2 (1-\beta)^2}
\Bigg )\ee 
where
$$
P_1^A(\beta,\om)=-\frac 
{\om(2-29\beta+51\beta \om+34\beta^2-30\beta^2 \om-7\beta^3+9\beta^3
  \om+18 \om)}{6(1-\om)}, 
$$
$$
P_2^A(\beta,\om)=-\frac
{\om^3(30\beta^2-9\beta^3+7\beta^3 \om-34\beta^2
  \om-2\om-51\beta+29\beta \om-18)}{6(1-\om)}, 
$$
$$
P_1^V(\beta,\om)=-\frac {\om (-2-7\beta+17\beta \om+14\beta^2-10\beta^2
\om-5\beta^3+3\beta^3 \om+6\om)}{2(1-\om)},
$$
$$
P_2^V(\beta,\om)=-\frac {\om^3 (-6+2\om-17\beta+10\beta^2-3\beta^3+7\beta
\om-14\beta^2 \om+5\beta^3 \om)}{2(1-\om)}.
$$

Substituting this result to eq. (\ref {gluonmassint}) and performing
the integration over $\lambda$ one gets
$$
\eta _{A,V} (b) \sim \int \limits_{0}^{1} d\beta 
\Bigg [\frac {P_1^{A,V}(\beta,\om)}{(1-\beta)^2}
~~\Pi \Bigg ( \frac {(1-\beta)^2}{\beta} \Bigg ) -
\frac {P_2^{A,V}(\beta,\om)}{\om ^2(1-\beta)^2}
~~\Pi \Bigg ( \frac {\om ^2 (1-\beta)^2}{\beta} \Bigg ) \Bigg ],
$$
where $\Pi(s)$ is the vacuum polarization function:
\be
\Pi (s)=\left(1-\frac {2}{s}\right) 
 v\ln \left( \frac {v+1}{v-1} \right)
+\frac {4}{s} - \frac {5}{3}
\ee
with
$$
v=\sqrt {1+\frac {4}{s}}.
$$
Subsequent calculation of the one-parametric integrals is
``tedious but straightforward.''  

So far we have
discussed the calculation of the massive fermion bubbles in what can
be called on-shell or QED like normalization. The final results which
we would like to get are the ones with the $\overline {MS}$--coupling
constant $\alpha _s$ renormalized at the symmetric point 
$\mu =\sqrt {m_bm_c}$.
The transformation to this form amounts to performing a scale
transformation in the coupling constant. Finally, for the $b$--quark bubble
contribution to the zero recoil form factors one gets:
$$
\eta _A (b) =\frac {(3\om^4-7\om^3-7\om^2-25\om+36)R_1 }{48 \om ^4} 
+\frac {(5\om-18\om^2-34\om^4-21\om^5+36)R_2}{48 \om ^4 (1-\om )}
$$
$$
-\frac {(1+\om)\ln^2 (\om )}{4(1-\om)}
-\frac {(15-4\om+18\om^2-37\om^3)\ln(\om)}{24(1-\om)\om^2}
+\frac {72-33\om+104\om^2-143\om^3}{72 \om^2 (1-\om)}
-\frac {\pi ^2 (5-3\om)}{36(1-\om)}
$$
and
$$
\eta _V (b) =\frac {(\om^4+3\om^3+3\om^2-
19\om+12)R_1 }{16\om^4}
+\frac {(-7\om^5+12-22\om^2-9\om-6\om^4)R_2}{16\om ^4 (1-\om)} 
$$
$$
-\frac {(1+\om)\ln^2 (\om)}{4(1-\om)}
-\frac {(15-36\om+14\om^2-25\om^3)\ln(\om)}{24\om^2(1-\om)}
+\frac {24-51\om+62\om^2-35\om^3}{24\om^2(1-\om)}
-\frac {\pi^2(3-\om)}{12(1-\om)}.
$$

\section{Results}
\label{sec:results}

It is convenient to divide up the 
functions $\eta_{A,V}^{(2)}$ into parts proportional to
various SU(3) factors (an overall factor $C_F$ has been factored out):
\begin{eqnarray}
\eta_{A,V}^{(2)} = C_F \eta_{A,V}^F + (C_A -2 C_F)\eta_{A,V}^{AF}
+T_R N_L \eta_{A,V}^L +T_R  \eta_{A,V}^H.
\label{eq:colors}
\end{eqnarray}
For a general SU(N) group $C_A=N$; $C_F=(N^2-1)/2N$; $T_R=1/2$.  $N_L$
denotes the number of the light quark flavors whose masses can be
neglected.  The last term contains contributions of the massive quark
loops, with $b$ and $c$ quarks.  We neglect the top quark; its impact
is suppressed by a factor $\sim m_b^2/m_t^2$.  We  use the pole
masses of $b$ and $c$ quarks and express our results in terms of 
$\alpha _{{\overline{MS}}}(\sqrt {m_b m_c})$.

Among the eight coefficient functions in eq.~(\ref{eq:colors})
$\eta_{A,V}^L$ are already known exactly \cite{Neubert95beta}.  They
corresponds to our diagram (f) in fig.~1 with a massless fermion in
the loop; they read
\begin{eqnarray}
\eta_V^L &=& {1\over 24}\left[ {1 +\om \over 1 - \om} \ln(\om)+2
\right],
\nonumber\\
\eta_A^L &=& {5\over 24}\left[ {1+\om \over 1- \om} \ln(\om)+
{44\over 15}\right].
\end{eqnarray}

The contribution of the massive quark loops is obtained by combining
the result for the $b$ quark loop, obtained in the previous section,
with an analogous formula for the $c$ quark.  The latter is obtained
 from the former by a simple symmetry transformation.  We find
\be
\lefteqn{\eta_V^H =
  (1 - \omega)^3 (1 + \omega) {12 + 5 \omega + 13 \omega^2 + 5 \omega^3 + 
       12 \omega^4\over 16 \omega^4}  R_1}
\nonumber\\&&
+  (1 + \omega) {12 - 21 \omega - \omega^2 + \omega^3 - 14 \omega^4 +
  \omega^5 - \omega^6 - 21 \omega^7 + 12 \omega^8\over 
            16 (1 - \omega)\omega^4}R_2
\nonumber\\&&
+  \lnomegad {1 + \omega + 5 \omega^4 - 3 \omega^5\over 4 (1 - \omega)}
  -\lnomega (1 + \omega) {15 - 51 \omega + 40 \omega^2 - 51 \omega^3 + 
        15 \omega^4 \over 24 (1 - \omega) \omega^2}
\nonumber\\&&
  -{37 - 30 \omega + 66 \omega^3 - 53 \omega^4 + 12 \omega^5 \over
   96 (1 - \omega)}\pi^2
+  {24 - 27 \omega + 70 \omega^2 - 27 \omega^3 + 24 \omega^4\over 24
  \omega^2}, 
\\[2mm]
\lefteqn{\eta_A^H =
  (1 - \omega^2) {36 - 25 \omega + 29 \omega^2 - 32 \omega^3 + 
       29 \omega^4 - 25 \omega^5 + 36 \omega^6 \over 48 \omega^4}R_1}
\nonumber\\&&
 + (1 + \omega) {36 - 31 \omega + 13 \omega^2 - 13 \omega^3 - 42 \omega^4 - 
       13 \omega^5 + 13 \omega^6 - 31 \omega^7 + 36 \omega^8\over
   48 (1 - \omega) \omega^4}R_2
\nonumber\\&&
 + \lnomegad {3 + 3 \omega + 7 \omega^4 - 9 \omega^5\over 12 (1 - \omega)}
  -\lnomega (1 + \omega) {15 - 19 \omega - 19 \omega^3 + 15 \omega^4
   \over 24 (1 - \omega) \omega^2}
\nonumber\\&&
+  {72 + 39 \omega + 286 \omega^2 + 39 \omega^3 + 72 \omega^4
 \over 72 \omega^2}
  -{79 - 10 \omega + 54 \omega^3 - 127 \omega^4 + 36 \omega^5\over 
   288 (1 - \omega)}\pi^2.
\ee
The functions $R_{1,2}$ are defined in eq.~(\ref{eq:r12}).

\begin{figure} 
\hspace*{0mm}
\begin{minipage}{16.cm}
\vspace*{5mm}
\[
\mbox{
\begin{tabular}{cc} 
\psfig{figure=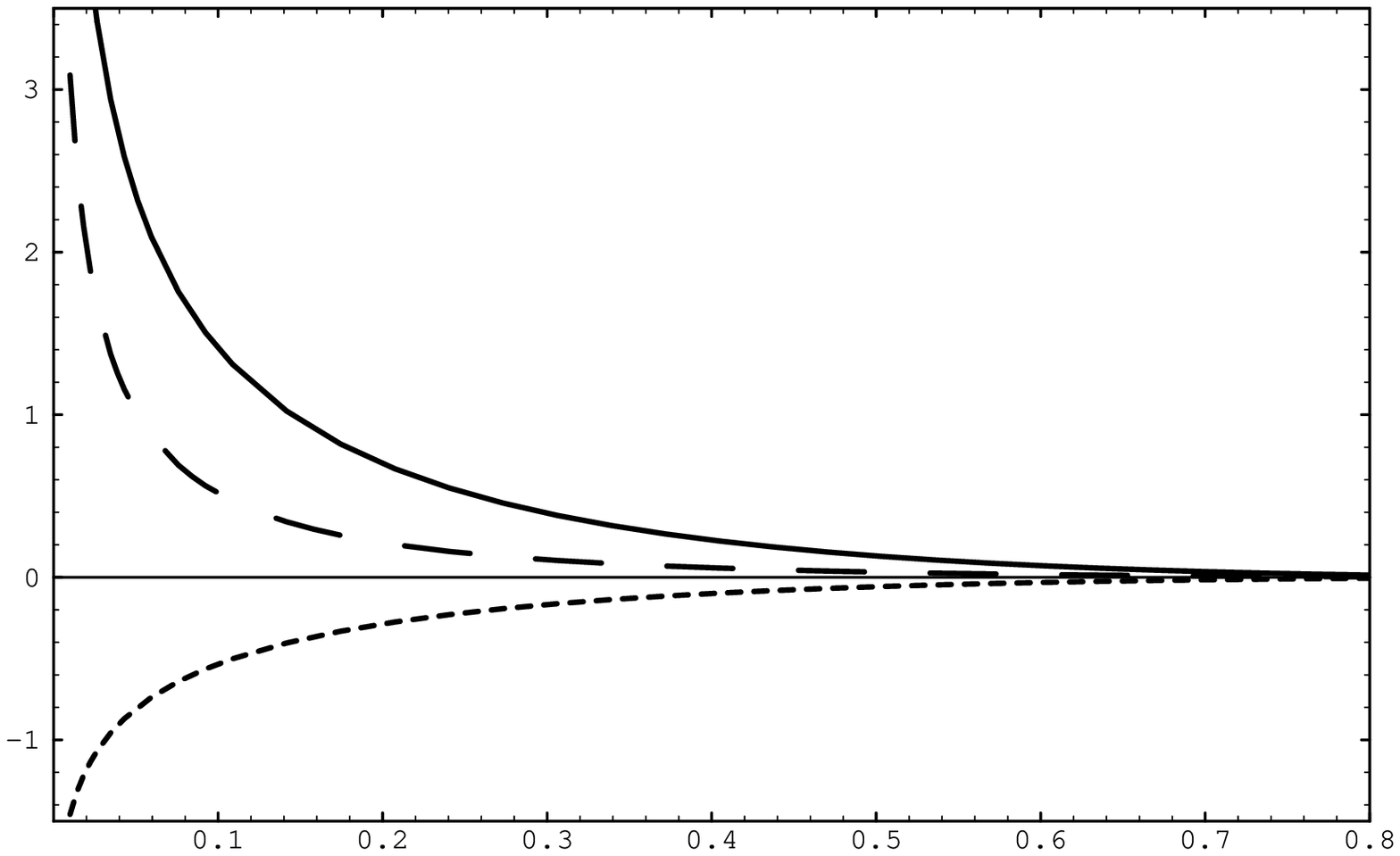,width=75mm,bbllx=72pt,bblly=270pt,%
bburx=540pt,bbury=520pt} 
&\hspace*{1mm}
\psfig{figure=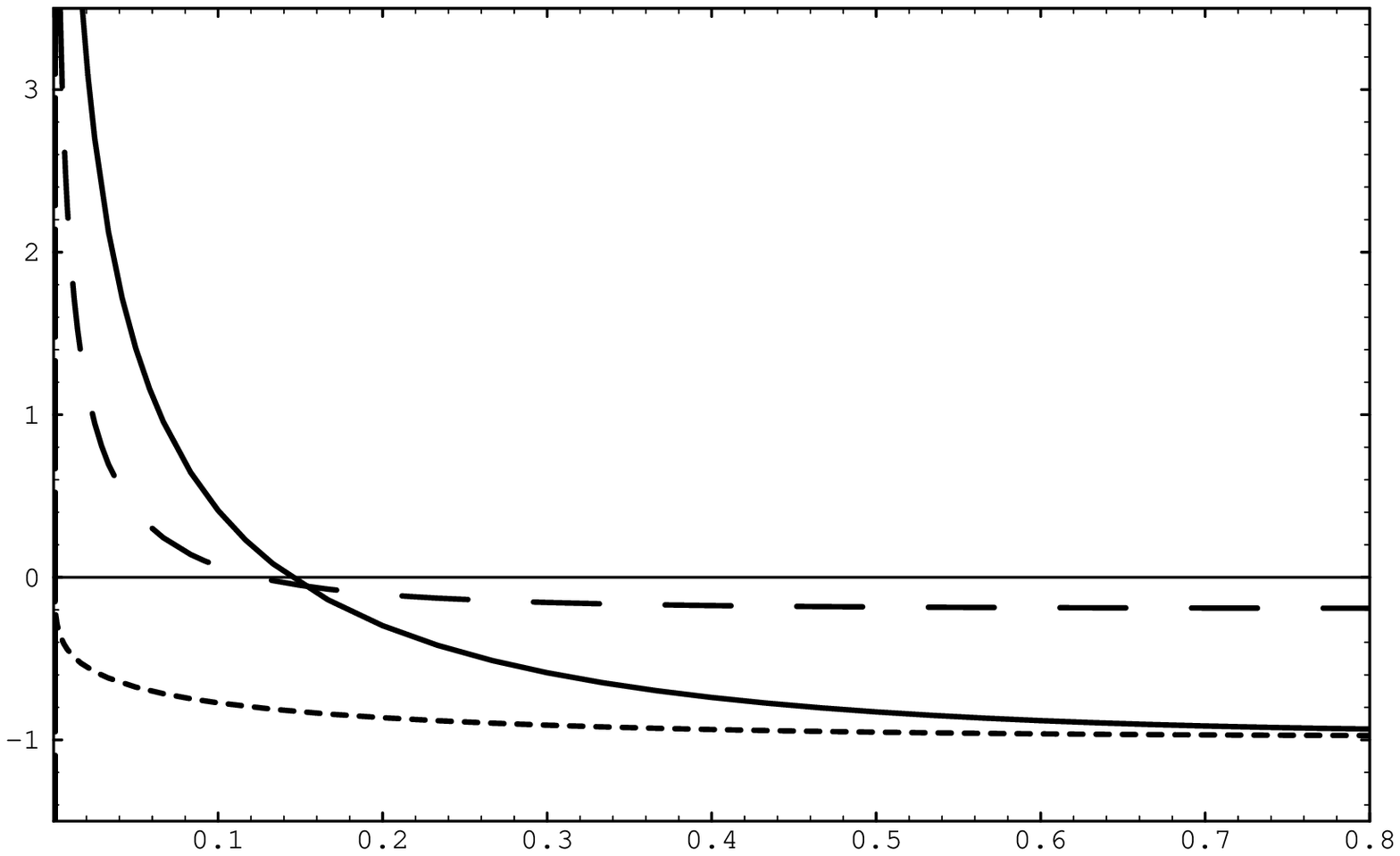,width=75mm,bbllx=72pt,bblly=270pt,%
bburx=540pt,bbury=520pt} 
\\[5mm]
\hspace*{-5mm}(a) & \hspace*{7mm}(b) 
\end{tabular}}
\]
\end{minipage}
\caption{Coefficient functions $\eta^F$ (solid), $\eta^{H}$ (dashed), 
$\eta^{AF}$ (dotted), for the vector (a) and axial case (b), plotted
against $\om = 0..0.8$; none of these functions varies noticably
between $\om = 0.8$ and $\om = 1$.}
\label{fig:plots}
\end{figure}

Diagrams without fermion loops give rise to two color structures and
we chose the combinations $C_F^2$ and $C_A C_F-2C_F^2$ to describe
them.  
For the functions $\eta_{V,A}^F$ we find
\be
\lefteqn{\eta_V^F =
 - {\omega^2 (2+\omega) \over 8(1-\omega)^2 } \lnomegad
 + \left({89\over 96}+ {\pi^2\over 6}\right)
       {1+\omega \over 1-\omega}\lnomega  
 + {(1+\omega)^3 \over 4 \omega (1-\omega)} \Li_2(1-\omega)
}
\nonumber \\ &&
       - {1-10\omega+\omega^2\over 24\omega}  \pi^2 
       + {53\over 48} 
 +{ 3-10 \omega+3 \omega^2 \over (1-\omega)^2} f(\omega),
\\[2mm]
\lefteqn{\eta_A^F =
  {\omega^2(6-\omega) \over 24(1-\omega)^2}\lnomegad 
 + \left({53\over 96}+{\pi^2\over 6}\right) 
         {1+\omega\over 1-\omega} \lnomega   
+{(1+\omega) (1-6 \omega+\omega^2)\over 
 12 \omega (1-\omega)} \Li_2(1-\omega)}
\nonumber \\ &&
 -{1-38\omega+\omega^2\over 72\omega} \pi^2
 - {95\over 72}
 +{9-14 \omega+9 \omega^2 \over  3(1-\omega)^2} f(\omega),
\\[2mm]
&&f(\omega)=
 \left( \Li_2(1-\omega) - {\pi^2\over 4}\right)  \lnomega 
 + {3\over 2} \left(\Li_3(\omega)  -\zeta_3\right) 
 + {3\over 4} \left(\lndelta+{1\over 8}\right) \lnomegad.
\nonumber
\ee
The most complicated and difficult to compute are the functions
 $\eta_{V,A}^{AF}$.  We parameterize them using, in addition to the
 notations introduced before, in terms of four functions $f_{1\ldots
 4}(\omega)$ for which closed
 formulas are given in the appendix.  In the following formulas we
 drop the argument $\omega$ of $f_i$:
\be
\lefteqn{   \eta_V^{AF} =
 {f_1 \omega \over 8}(1 - 3 \omega)
 -f_2 (1 + \omega) {1 + 2 \omega - \omega^2 \over 16 \omega}
 -f_3 {1 - \omega + 5 \omega^2 - \omega^3 \over 16 (1 - \omega)}
 + f_4 {\omega (1 + \omega) \over 8 (1 - \omega)}}
\nonumber \\ &&
 -{(1 - \omega^2) (3 R_1 + R_2) \over 16 \omega}
       - R_1 \lnomega   
       + {3\over 2}\left(\Li_3(\omega) -\zeta_3\right)
       + {3\over 4}\lndelta \lnomegad
       -{1\over 8} \omega \lnomegad
\nonumber \\ &&
-{17\over 96} \lnomega {1 + \omega \over 1 - \omega}
-\lnomega {(1 + \omega)^2 \pi^2 \over 24 (1 - \omega) \omega}
        + {5\over 96}(1- \omega) \pi^2 
       - {17\over 48 },
\label{eq:vf}
\\[2mm]
\lefteqn{   \eta_A^{AF} =
       - R_1 \lnomega
       +{3\over 2} \Li_3(\omega)  
+  {(1 - \omega^2) (8 - \omega + 8 \omega^2) \over 48 \omega^2} R_1}
\nonumber \\ &&
+  {(1 + \omega) (8 - 3 \omega + 22 \omega^2 - 3 \omega^3 + 8 \omega^4)\over
   48 (1 - \omega) \omega^2} R_2
      +{3\over 4} \lndelta \lnomegad
 -{f_1 \omega\over 24}  (1 + 9 \omega)
\nonumber \\ &&
- f_2 {3 + \omega + 19 \omega^2 - 3 \omega^3 \over 48 \omega}
 -f_3 {3 - 11 \omega + 31 \omega^2 - 3 \omega^3 \over 48 (1 - \omega)}
+ f_4 {\omega (7 + 3 \omega) \over 24 (1 - \omega)}
\nonumber \\ &&
  -\lnomegad \omega {9 - 6 \omega + 5 \omega^2 - 4 \omega^3 \over 24
 (1 - \omega)^2} 
 -\pi^2 \lnomega {3 - 2 \omega + 3 \omega^2 \over 72 (1 - \omega) \omega}
 -{61\over 96} \lnomega {1 + \omega \over  1 - \omega}
\nonumber \\ &&
+{63 - 46 \omega + 23 \omega^2 - 8 \omega^3 \over 288 (1 - \omega)}\pi^2
       - {151\over 72} - {\pi^2 \ln 2\over 6}  - {5\over 4} \zeta_3.
\label{eq:af}
\ee
The behavior of the  above coefficient functions is plotted in
Fig.~\ref{fig:plots}.  It can be seen that the corrections to the
vector coupling vanish when $\om\to 1$, in accordance with vector
current conservation.
For  $\omega=0.3$, which roughly corresponds to the physical value of
the mass ratios of $c$ and $b$ quarks, we find the following results 
\be
&& \eta_V^H = 0.107667, \quad \eta_V^F = 0.393822, 
                        \quad \eta_V^{AF} = -0.167695, 
\nonumber \\
&& \eta_A^H = -0.154818, \quad \eta_A^F = -0.586889,
                        \quad \eta_A^{AF} = -0.909398,
\ee
in excellent agreement with the approximate values found in
\cite{zerorecoil} (see eq.~(11) in that reference). 

From the above exact formulas we can also find the expansions of all
the coefficient functions around $\om=0$ which corresponds to the
vanishing $c$ quark mass.  We give below the terms which do not vanish
as $\om\to 0$:
\be
\eta_V^H &\to &    {\lnomegad\over 4}  
 + {13\lnomega\over 24} +{1147\over 288} - {37\pi^2\over 96}, 
\nonumber \\
\eta_A^H &\to & 
   {\lnomegad\over 4} + {7\,\lnomega\over 8}  
+ {971\over 288}  - {79\,\pi ^2\over {288}}, 
\nonumber \\
\eta_V^F &\to &    {9\,\lnomegad\over 32}
+\left( {113\over 96}-{\pi^2\over 12}\right)\lnomega
+  {41\over 48}
 + {7\,\pi^2\over 12} - {9\,\zeta_3\over 2},
\nonumber \\
\eta_A^F &\to &    {9\,\lnomegad\over 32}
+\left({61\over 96}-{\pi^2\over 12}\right)\lnomega
  -{101\over 72}  + {17\,\pi ^2\over 36} - {9\,\zeta_3\over 2},
\nonumber \\
\eta_V^{AF} &\to &  \left(- {5\over 96}+ {\pi^2\over 24}\right)\lnomega
  -{23\over 48} - 
   {\pi^2\over 96}  + 
   {\pi ^2\,\ln(2)\over 2} - {9\,\zeta_3\over 4},
\nonumber \\
\eta_A^{AF} &\to & 
 \left( - {49\over 96}+ {\pi ^2\over 24}\right)\lnomega
  -{83\over 36} +    {5\,\pi ^2\over 32}  + 
   {\pi^2\,\ln (2)\over 3} - 2\,\zeta_3.
\ee

\section{Conclusions}
In this paper we have presented an analytical calculation 
of the ${\cal O}(\alpha _s ^2)$ corrections to the functions $\eta _A$
and $\eta _V$ which parameterize semileptonic $b \to c$ transition at
zero recoil. Numerically, our results are in excellent agreement
with the results of the ref.~\cite {zerorecoil} obtained using an
approximate approach to the present problem.

This calculation was facilitated by 
an efficient method of
computing two-loop vertex functions with two mass scales in the zero
recoil limit which we  demonstrated in some detail.  
Infrared singularities of individual diagrams 
were shown to be the same as in
certain combinations of single scale integrals, for which an exact
algorithm is known.  We presented various useful tricks which permit a
calculation of the remaining two-scale integrals. Since the
previous approximate formulas \cite {zerorecoil}
were valid only for similar masses of
initial and final quarks, the exact formulas we obtained significantly
extend our knowledge of two-loop corrections to fermion decays.

\section*{Acknowledgement}
This research was supported by the grant BMBF 057KA92P and by 
``Gra\-duier\-ten\-kolleg
Elementarteilchenphysik'' at the University of Karlsruhe.

\appendix
\section{Analytical results for the functions $f_i$}
In eq.~(\ref{eq:vf}-\ref{eq:af}) we presented the results for the
coefficients 
$\eta^{AF}$ in terms of four functions $f_{1\ldots 4}$.  
In this
appendix we present closed formulas for $f_1$ and $f_4$.  
The remaining two functions $f_{2,3}$ are obtained from these
results by applying the following symmetry relations:
$$
f_2(\om)=\frac {1}{\om^2} f_1 \Bigg ( \frac {1}{\om} \Bigg ),
$$
$$
f_3(\om) = \frac {1}{\om ^2} f_4 \Bigg ( \frac {1}{\om} \Bigg ).
$$

For $f_1(\om)\equiv D(0,1,0,1,0,1,0,1,1)$ we find
$$
\frac{1}{\om(1-\om)} \Bigg [
4\Li _3(\om)+8\Li _3 \left(\frac {1}{1+\om}\right)
+4\Li _3(1-\om)
+6\Li _3(-\om)
-2\Li _3(1-\om^2)
$$
$$
-4\Li _3 \left(\frac {1+\om}{2} \right)
+4\Li _3 \left( \frac {2 \om}{1+\om} \right)
-\frac {13}{2} \zeta _3
-4\Li _2 \left( \frac {2\om}{1+\om} \right)\ln(\om)
-6\ln(\om)\Li _2(-\om)
-2\ln(\om)\Li _2(\om)
$$
$$
+2\ln(2)\ln^2(1+\om)
-2\ln^2(2)\ln(1+\om)
-2\ln^2(\om)\ln(1-\om)
-\ln^2(\om)\ln(1+\om)
$$
$$
+ \pi^2\ln(1+\om)
-\frac {\pi^2}{3}\ln(2)
+\frac {2}{3}\ln^3(2)
-2\ln^3(1+\om) \Bigg ].
$$

The result for $f_4(\om)\equiv D(1,1,0,1,0,1,0,1,0)$ is
$$
\frac {1}{\om} \Bigg [ 
-2\Li _3(1-\om^2)
-4\Li _3 \left( \frac {1+\om}{2} \right)
+4\Li _3(1-\om)
+8\Li _3 \left( \frac {1}{1+\om} \right)
-2\Li _3(\om)
$$
$$
+4\Li _3 \left(\frac {2\om}{1+\om} \right)
-\frac {13}{2}\zeta _3
-4\Li _2 \left(\frac {2\om}{1+\om} \right)\ln(\om)
+2\ln(\om)\Li _2(\om)
-2\ln(\om)\Li _2(-\om)
$$
$$
+2\ln(2)\ln^2(1+\om)
-\frac {\pi^2 }{3} \ln (2)
+2\pi ^2\ln(1+\om)
-\ln^2(\om)\ln(1-\om)
+\frac {2}{3}\ln^3(2)
$$
$$
-2\ln^3(1+\om)
-2\ln^2(2)\ln(1+\om)
\Bigg ]. 
$$


\end{document}